\documentclass[aps,prl,preprint,superscriptaddress]{revtex4-1}
\usepackage{graphicx}
\begin{document}

%Title of paper
\title{Presence of magnetic excitations in SmFeAsO}

\author{Jonathan Pelliciari}
\email[]{jonathan.pelliciari@psi.ch}
\affiliation{Swiss Light Source, Paul Scherrer Institut, CH-5232 Villigen PSI, Switzerland}
\author{Marcus Dantz}
\affiliation{Swiss Light Source, Paul Scherrer Institut, CH-5232 Villigen PSI, Switzerland}
\author{Yaobo Huang}
\affiliation{Swiss Light Source, Paul Scherrer Institut, CH-5232 Villigen PSI, Switzerland}
\affiliation{Beijing National Lab for Condensed Matter Physics, Institute of Physics, Chinese Academy of Sciences P. O. Box 603, Beijing 100190, China}
\author{Vladimir N. Strocov}
\affiliation{Swiss Light Source, Paul Scherrer Institut, CH-5232 Villigen PSI, Switzerland}
\author{Lingyi Xing}
\affiliation{Beijing National Lab for Condensed Matter Physics, Institute of Physics, Chinese Academy of Sciences P. O. Box 603, Beijing 100190, China}
\author{Xiancheng Wang}
\affiliation{Beijing National Lab for Condensed Matter Physics, Institute of Physics, Chinese Academy of Sciences P. O. Box 603, Beijing 100190, China}
\author{Changqing Jin}
\affiliation{Beijing National Lab for Condensed Matter Physics, Institute of Physics, Chinese Academy of Sciences P. O. Box 603, Beijing 100190, China}
\affiliation{Collaborative Innovation Center for Quantum Matters, Beijing 100871, China}
\author{Thorsten Schmitt}
\email[]{thorsten.schmitt@psi.ch}
\affiliation{Swiss Light Source, Paul Scherrer Institut, CH-5232 Villigen PSI, Switzerland}

\date{\today}

\begin{abstract}
We measured dispersive spin excitations in $\mathrm{SmFeAsO}$, parent compound of $\mathrm{SmFeAsO_{\text{1-x}}F_{\text{x}}}$ one of the highest temperature superconductors of Fe pnictides (T$_{\text{C}}\approx$55~K). We determine the magnetic excitations to disperse with a bandwidth energy of ca 170 meV at (0.47, 0) and (0.34, 0.34), which merges into the elastic line approaching the $\Gamma$ point. Comparing our results with other parent Fe pnictides, we show the importance of structural parameters for the magnetic excitation spectrum, with small modifications of the tetrahedron angles and As height strongly affecting the magnetism.
\end{abstract}

\maketitle
Since the discovery of high temperature superconductivity (SC) \cite{kamihara_iron-based_2008} in $\mathrm{LaFeAsO_{1-x}F_x}$ the number of Fe-based compounds has quickly boosted and new families differing in structure and stoichiometry have been discovered and synthesized, the most common being the $\mathrm{SmFeAsO}$ (1111), $\mathrm{BaFe_2As_2}$ (122), $\mathrm{NaFeAs}$ (111), and $\mathrm{FeSe}$ (11) (see \cite{stewart_superconductivity_2011, johnston_puzzle_2010} for extended reviews). Ubiquitous FeAs layers, composed of FeAs$_4$ tetrahedrons, are separated by spacers, which differ from family to family. Generally, the parent compounds are antiferromagnetically ordered, and SC emerges upon hole, electron or isovalent doping, with the dopants either in the FeAs or in the spacing layer \cite{stewart_superconductivity_2011, johnston_puzzle_2010}. The phase diagram, characterized by antiferromagnetism and SC, is similar to other unconventional superconductors, such as the cuprates and heavy fermions systems where a magnetic-mediated superconducting pairing mechanism has been proposed \cite{scalapino_common_2012, chubukov_pairing_2012}. Similarly, such a scenario has been further extended from these systems to Fe pnictides \cite{scalapino_common_2012, chubukov_pairing_2012}. In this framework, residual antiferromagnetic (AF) fluctuations are expected to be strong and possibly lead to a superconducting phase. Moreover, the structural / nematic transition at T$>$T$_N$ has been associated to spin excitations \cite{fernandes_what_2014,stewart_superconductivity_2011,johnston_puzzle_2010}. Thus, the detection of these AF fluctuations within several families with different structures is of vital importance for a complete understanding of Fe pnictides.

\begin{figure}
\includegraphics[scale=0.4]{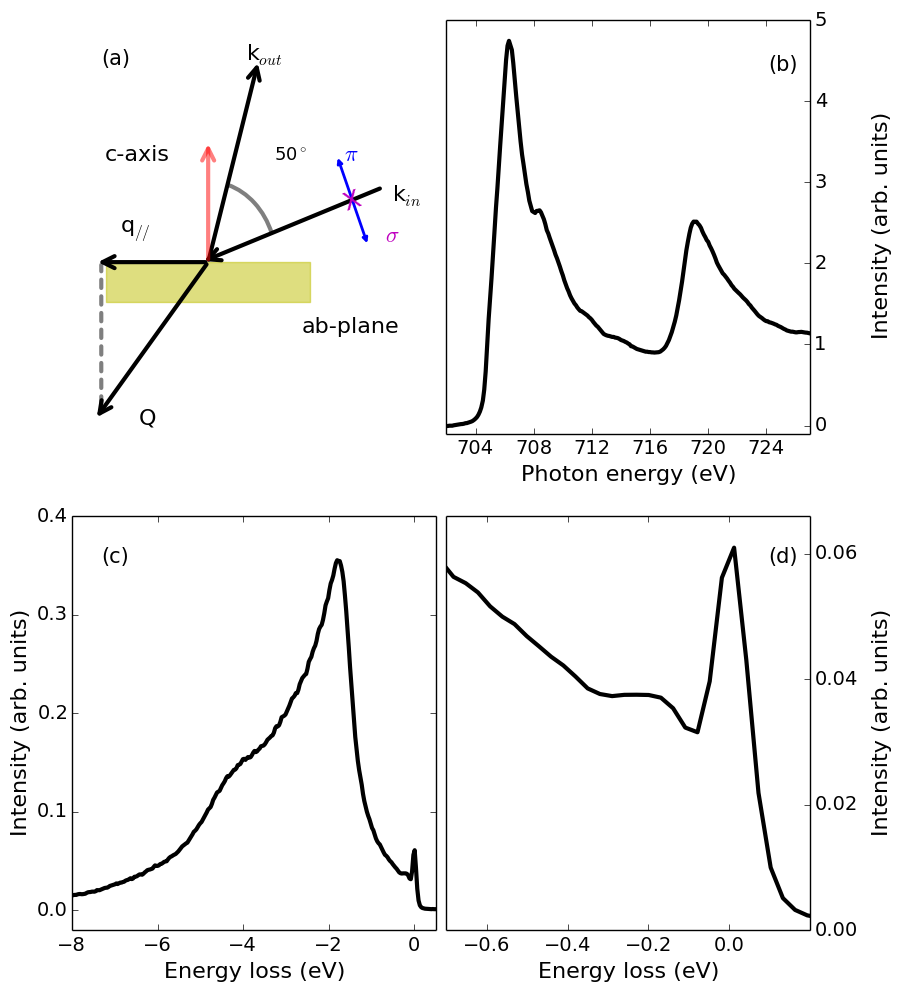}
\caption{\label{fig:fig1} (a) Experimental geometry of RIXS experiments. (b) XAS spectrum of $\mathrm{SmFeAsO}$ at 15$^{\circ}$ between incoming beam and sample surface along (0, 0)$\rightarrow$(1, 0) with $\pi$ polarization. (c) RIXS spectrum of $\mathrm{SmFeAsO}$ at 15$^{\circ}$ between incoming beam and sample surface along (0, 0)$\rightarrow$(1, 0). Polarization was set to $\pi$ and the incident energy was tuned on the maximum of the XAS spectrum shown in (b). (d) Zoom into the low energy region of (c). Temperature is 10~K for all the spectra.}
\end{figure}

The effect of magnetism on the electronic structure has been measured by Angle Resolved Photo-Emission Spectroscopy (ARPES) through the detection of a kink in the band structure \cite{richard_angle-resolved_2009,liu_unusual_2010,yang_electronic_2011}, which has been ascribed to electron-boson coupling with the bosonic candidate being of magnetic nature. However, ARPES represents an indirect spectroscopy to characterize magnetism which has to be characterized by means of techniques sensitive to spin excitations, such as neutron scattering \cite{dai_antiferromagnetic_2015,tranquada_superconductivity_2014,fujita_progress_2011,inosov_spin_????,zhang_effect_2014} and / or X-Ray Scattering \cite{zhou_persistent_2013, pelliciari_intralayer_2016}. Neutron scattering experiments confirmed the presence of sizable magnetic moments in Fe pnictides (on the order of $\approx$ 1 $\mu_B$) \cite{dai_antiferromagnetic_2015,johnston_puzzle_2010,stewart_superconductivity_2011} with few exceptions, such as $\mathrm{NaFeAs}$ that shows lower ordered magnetic moment \cite{johnston_puzzle_2010,stewart_superconductivity_2011}. On the dynamical side, spin wave-like excitations were also observed in the AF phases of several compounds by inelastic neutron scattering (INS) and Resonant Inelastic X-Ray Scattering (RIXS) \cite{dai_antiferromagnetic_2015,luo_electron_2012,luo_electron_2013,pelliciari_intralayer_2016,wang_doping_2013,zhang_effect_2014,zhou_persistent_2013}. The importance of spin fluctuations has been further confirmed by their persistence within the superconducting phase \cite{zhou_persistent_2013,wang_doping_2013, pelliciari_intralayer_2016,dai_antiferromagnetic_2015,luo_electron_2012,luo_electron_2013}, even though a conclusive picture of their role is still under development.

\begin{figure}
\includegraphics[scale=0.4]{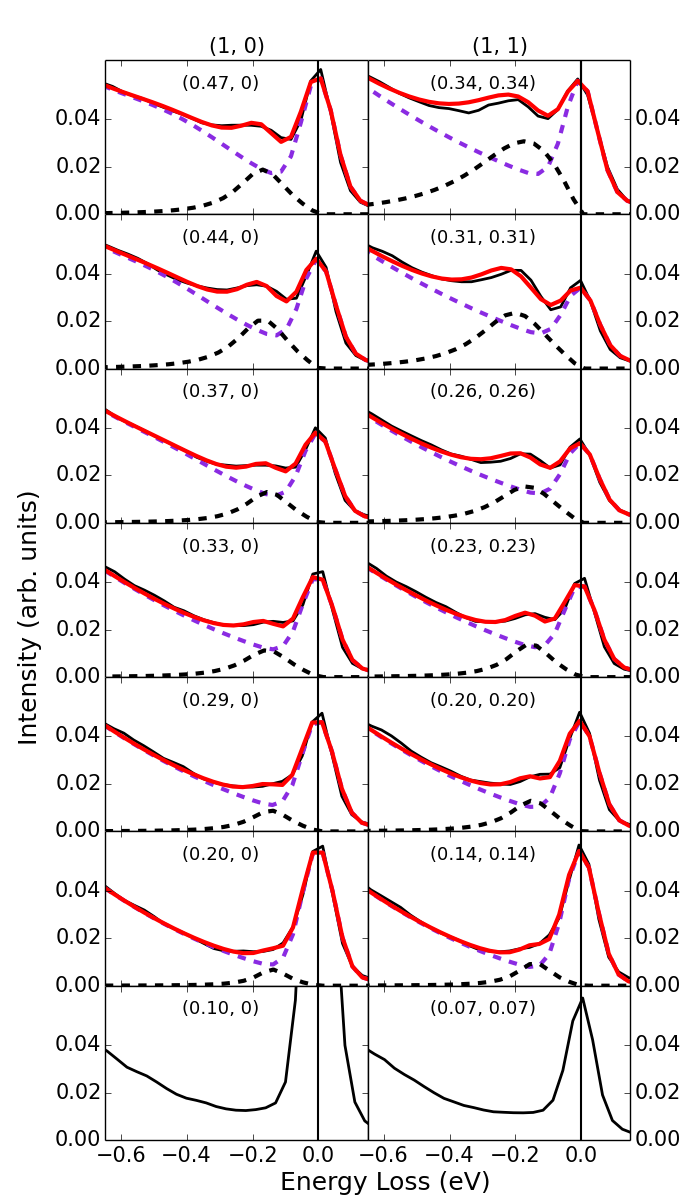}
\caption{\label{fig:fig2} Momentum dependence of RIXS spectra along (0, 0)$\rightarrow$(0.47, 0) and (0, 0)$\rightarrow$(0.34, 0.34 r.l.u.) for $\mathrm{SmFeAsO}$. Spectra were recorded in $\pi$ polarization at the maximum of the Fe L$_{3}$ absorption edge at 10~K. Experimental data are shown as black solid line, fitting of background and elastic line are the purple dashed line, and the magnetic peaks are depicted as black dashed line for every spectrum. The total fitting is depicted as red solid line. At low q$_{//}$ a fitting is unreliable, so no fitting is attempted.}
\end{figure}

1111 crystals are known for naturally cleaving polarly which makes the interpretation of surface sensitive spectroscopic data (such as ARPES) difficult because of the mixing of surface states with bulk states \cite{de_jong_high-resolution_2009,eschrig_calculated_2010,yang_surface_2010,liu_surface-driven_2010,yang_electronic_2011}. Moreover, the growth of suitable crystals for INS is challenging and this complicates the measurements of high energy spin excitations ($>$ 90 meV), where INS would provide plenty of information \cite{dai_antiferromagnetic_2015, tranquada_superconductivity_2014, fujita_progress_2011, ramazanoglu_two-dimensional_2013}. These drawbacks can be minimized by employing RIXS, which has been previously employed in the detection of high energy spin excitations in $\mathrm{NaFeAs}$ and $\mathrm{BaFe_2As_2}$ Fe pnictides \cite{zhou_persistent_2013, pelliciari_intralayer_2016}. Moreover, thanks to the high refocusing (beam spot of 5x20 $\mu$m$^2$ VxH at the ADRESS beamline of the Swiss Light Source) and flux obtainable at the sample for modern beamlines \cite{strocov_high-resolution_2010}, the amount of sample required for these investigations is on the order of tens of milligrams and crystals of the size 150x200 micrometers can now be successfully studied, even measuring down to a single layer of material \cite{dean_spin_2012}.

In this Letter, we report on the measurement of high energy spin excitations in $\mathrm{SmFeAsO}$, a parent compound of the 1111 series. We identify dispersing magnetic excitations ranging up to an energy of 170 meV at (0.47, 0 r.l.u.) which merge into the elastic line at the $\Gamma$ point. Similar behavior is detected along the diagonal direction, where spin excitations disperse to 160-170 meV at (0.34, 0.34 r.l.u.) and decrease in energy moving towards the $\Gamma$ point. The spin excitations bandwidth has a value similar to $\mathrm{BaFe_2As_2}$ ($\approx$190 meV) but higher than $\mathrm{NaFeAs}$ ($\approx$150 meV). We correlate the bandwidth of magnetic excitations with the structure of the $\mathrm{FeAs_4}$ tetrahedron, in particular the height of As respect to the Fe layer (h$_{FeAs}$). The detection of high energy spin excitations in this parent compound confirms that magnetic excitations are universal within the parent Fe pnictides.

Single crystals of $\mathrm{SmFeAsO}$ have been grown by the flux method, using $\mathrm{NaAs}$, $\mathrm{SmAs}$, $\mathrm{Fe_2O_3}$, $\mathrm{Fe}$ and $\mathrm{As}$ powder as starting materials. The precursor $\mathrm{NaAs}$ has been obtained by mixing $\mathrm{Na}$ lump and $\mathrm{As}$ powder, which had been sealed in an evacuated titanium tube and sintered at 650$^{\circ}$~C for 10 h. $\mathrm{SmAs}$ has been prepared by mixing $\mathrm{Sm}$ pieces and $\mathrm{As}$ powder, sealed in a evacuated $\mathrm{Ti}$ tube, and sintered at 700$^{\circ}$~C for 20 h. The stoichiometric amount of $\mathrm{NaAs}$, $\mathrm{SmAs}$, $\mathrm{Fe_2O_3}$, $\mathrm{Fe}$, and $\mathrm{As}$ powder have been weighed to achieve an element ratio of $\mathrm{NaAs}$ : $\mathrm{SmFeAsO}$ = 20 : 1. The mixture has been grounded thoroughly and put into an alumina crucible and sealed in an $\mathrm{Nb}$ crucible under 1 atm of Argon gas, which was then sealed in an evacuated quartz tube. Finally the mixture was heated to 1100$^{\circ}$~C and cooled slowly down to 700$^{\circ}$~C at a rate of 5$^{\circ}$~C / h to grow the single crystals. 

The samples were mounted with the \textit{ab} plane perpendicular to the scattering plane and the \textit{c} axis lying in it (sketch in Fig.~\ref{fig:fig1}(a) and post-cleaved \textit{in situ} at a pressure better than 2.0x10$^{-10}$ mbar. The directions studied are (1, 0) and (1, 1) according to the orthorhombic unfolded crystallographic notation \cite{park_symmetry_2010}. We use the convention of 1 Fe per unit cell. All the measurements were carried out at 10~K.
X-ray Absorption Spectra (XAS) and RIXS experiments were performed at the ADRESS beamline of the Swiss Light Source, Paul Scherrer Institute, Villigen, Switzerland \cite{strocov_high-resolution_2010, ghiringhelli_saxes_2006}. XAS spectra were measured in Total Fluorescence Yield (TFY). The RIXS spectrometer was set to a scattering angle of 130$^{\circ}$ and the incidence angle on the samples surface was varied to change the in plane momentum transferred (q) from (0, 0) to (0.47, 0 r.l.u.) (relative lattice units expressed in q$_{//}\cdot$a/2$\pi$) and from (0, 0) to (0.34, 0.34 r.l.u.) as shown in Fig.~\ref{fig:fig1}(a). All measurements are recorded in grazing incidence configuration. The total energy resolution was 110 meV, measured by means of elastic scattering from a carbon-filled acrylic tape.

We measured Fe L$_{2, 3}$ XAS spectra for the two crystallographic orientations at 15$^{\circ}$ incidence angle and $\pi$ polarization. Fig.~\ref{fig:fig1}(b) displays the XAS spectrum of the (1, 0) orientation. The XAS along (0, 0)$\rightarrow$(1, 1) (not displayed) is analogue to the (0, 0)$\rightarrow$(1,0) direction. The spectrum is composed of a broad peak centered at 707 eV, typical of metallic systems containing Fe \cite{zhou_persistent_2013, kurmaev_identifying_2009, yang_evidence_2009,pelliciari_intralayer_2016}. The incident energy for RIXS was tuned at the main Fe-L$_3$ peak. In Fig.~\ref{fig:fig1}(c) an exemplary RIXS spectrum of $\mathrm{SmFeAsO}$ at (0.47, 0) is shown. The main line in this spectrum resembles emission from metallic systems with a broad asymmetric peak displaying a maximum at around -2 eV in energy loss ($\hbar\nu_{out}$-$\hbar\nu_{in}$) \cite{zhou_persistent_2013, kurmaev_identifying_2009, hancock_evidence_2010,pelliciari_intralayer_2016}, arising from resonant emission of itinerant electrons. In contrast to doped 1111 \cite{nomura_resonant_2016}, but in agreement with other Fe pnictides, the emission line is not showing any sharp features ascribable to dd-excitations \cite{zhou_persistent_2013, kurmaev_identifying_2009, yang_evidence_2009,pelliciari_intralayer_2016}. This confirms the moderate electronic correlations of Fe pnictides contrasting with Fe chalcogenides where dd-excitations have been observed employing hard X-ray RIXS \cite{gretarsson_resonant_2015}.

\begin{figure}
\includegraphics[width=\textwidth]{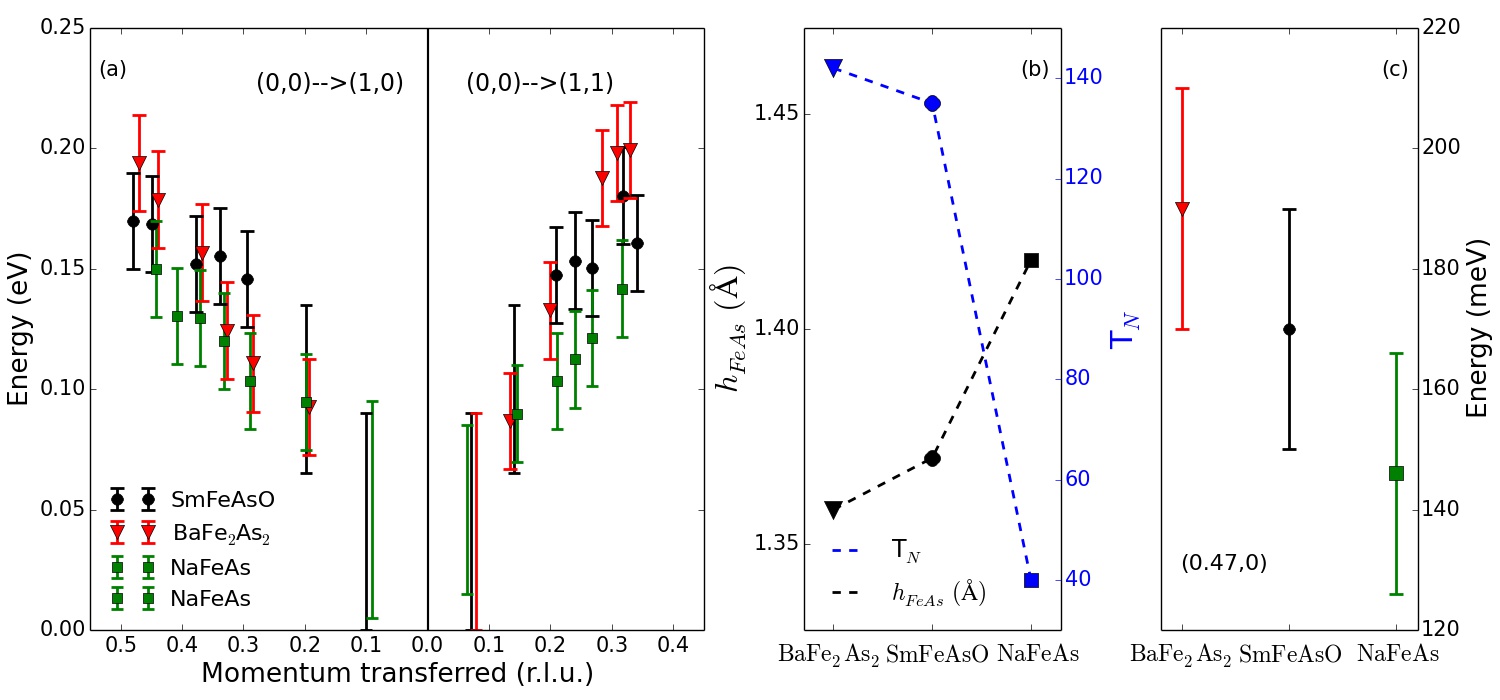}
\caption{\label{fig:fig3} (a) Dispersion of spin excitations in $\mathrm{BaFe_2As_2}$ \cite{zhou_persistent_2013}, $\mathrm{SmFeAsO}$, and $\mathrm{NaFeAs}$ detected along (0, 0)$\rightarrow$(1, 0) and (0, 0)$\rightarrow$(1, 1) extracted from RIXS experiments at 10~K. Close to (0, 0), we show only the error bar without marker because the peak of magnetic excitations and the elastic line are overlapped and a determination of the energy is unreliable (see text). (b) Comparison of T$_N$ and h$_{FeAs}$ in $\mathrm{BaFe_2As_2}$, $\mathrm{SmFeAsO}$, and $\mathrm{NaFeAs}$ from \cite{johnston_puzzle_2010,stewart_superconductivity_2011}. (c) Summary of the spin excitations energy at (0.47, 0).}
\end{figure}

On the low energy side of the RIXS spectra at (0.47, 0) and (0.34, 0.34), we observe a peak emerging from the background having an energy of 170 meV as illustrated in Fig.~\ref{fig:fig1}(d) and~\ref{fig:fig2}. This peak moves towards the elastic line when the in-plane momentum transferred is decreased and at low q$_{//}$ it merges into the elastic line. The energy range and the dispersive nature of this mode resemble the behavior of spin excitations detected by RIXS in $\mathrm{BaFe_2As_2}$ \cite{zhou_persistent_2013} and $\mathrm{NaFeAs}$ (measurements shown in Supplemental Material). We fit the background, the elastic and the magnetic line in agreement to \cite{pelliciari_intralayer_2016, zhou_persistent_2013, hancock_evidence_2010} and plot the results in Fig.~\ref{fig:fig2}. At low values of q$_{//}$, we do not attempt any fitting procedure because of the high overlap between the elastic and the magnetic peak. However, we believe that an estimation of a rough energy range is still possible. In Fig.~\ref{fig:fig3}(a) we show the dispersion curve arising from this fitting procedure as black dots with error bars together with the dispersion curves extracted from $\mathrm{BaFe_2As_2}$ \cite{zhou_persistent_2013} and $\mathrm{NaFeAs}$ measured in the same experimental conditions (see the raw data and the fittings of $\mathrm{NaFeAs}$ in the Supplemental Material). 

Figure~\ref{fig:fig3}(a) illustrates the dispersion relation of magnetic excitations for $\mathrm{SmFeAsO}$ (black dots with error bars), $\mathrm{NaFeAs}$ (green dots with error bars), and $\mathrm{BaFe_2As_2}$ (red dots with error bars) along (0, 0)$\rightarrow$(1, 0) and (0, 0)$\rightarrow$(1, 1). Clearly, there is a renormalization of the magnetic bandwidth between $\mathrm{BaFe_2As_2}$, that shows the highest energy (190 meV $\pm$20 meV), $\mathrm{SmFeAsO}$ displaying an intermediate energy (170 meV $\pm$20 meV), and $\mathrm{NaFeAs}$ having the lowest bandwidth (150 meV $\pm$20 meV) as summarized in Fig.~\ref{fig:fig3}(c). This trend is in qualitative agreement with the decreasing values of magnetic moment from $\mathrm{BaFe_2As_2}$ to $\mathrm{SmFeAsO}$ and $\mathrm{NaFeAs}$ \cite{stewart_superconductivity_2011,johnson_iron-based_2015,johnston_puzzle_2010,zhang_effect_2014,dai_antiferromagnetic_2015}. However a quantitative comparison between ordered magnetic moment and spin excitations is not straightforward and is beyond the goal of this paper.
The relevance of the structural parameters for SC and magnetism has been widely discussed in Refs.~\cite{stewart_superconductivity_2011,johnston_puzzle_2010,zhang_effect_2014}, with h$_{FeAs}$ and the angle of the tetrahedron formed by FeAs$_4$ as the main possible parameters affecting T$_C$ and T$_N$. Since there is no SC in these parent compounds, in Fig.~\ref{fig:fig3}(b) we show only the values of T$_{N}$ and how they correlate with h$_{FeAs}$ \cite{stewart_superconductivity_2011,johnston_puzzle_2010,zhang_effect_2014}. In details, h$_{FeAs}$ increases from 1.358 \AA~in $\mathrm{BaFe_2As_2}$ to 1.37 \AA~in $\mathrm{SmFeAsO}$ and then to 1.416 \AA~in $\mathrm{NaFeAs}$, whereas T$_N$ decreases \cite{stewart_superconductivity_2011,johnson_iron-based_2015,johnston_puzzle_2010,zhang_effect_2014,dai_antiferromagnetic_2015} as well as the magnetic bandwidth measured in our experiments and outlined in Fig.~\ref{fig:fig3}(c). This highlights the importance of the structure for the spin excitations in Fe pnictides as well as the expected role of structural deformations due to doping, pressure or even defects in phase transitions like nematicity and / or SC.

In conclusion, we measured the spin excitations spectrum of $\mathrm{SmFeAsO}$, parent compound of the 1111 series, bypassing the polar cleaving problem. Comparison with other parent Fe pnictides, measured in the same experimental configuration, show that the bandwidth of spin waves is slightly renormalized to lower values in $\mathrm{SmFeAsO}$ compared to $\mathrm{BaFe_2As_2}$ but higher than $\mathrm{NaFeAs}$. Here, we have illustrated how the structure, in particular h$_{FeAs}$, influences the magnetic properties of Fe pnictides possibly triggering most of the modifications. We claim that, if such structural modifications can affect the magnetism and bandwidth, then the perturbations provided by doping and / or pressure might lead to instabilities such as Cooper pairing and / or nematicity, not triggered by electronic effects alone but by structural effects as well.

%\section{Supplementary Material}
%See supplementary materials for the raw RIXS data of $\mathrm{NaFeAs}$.

\begin{acknowledgments}
J.P. and T.S. acknowledge financial support through the Dysenos AG by Kabelwerke Brugg AG Holding, Fachhochschule Nordwestschweiz, and the Paul Scherrer Institut. Part of this research has been funded by the Swiss National Science Foundation through the D-A-CH programme (SNSF Research Grant 200021L 141325). Experiments have been performed at the ADRESS beamline of the Swiss Light Source at Paul Scherrer Institut. The work at IOP-CAS is supported by NSF and MOST through research projects.
\end{acknowledgments}

%\bibliography{Zotero.bib}{}
%merlin.mbs apsrev4-1.bst 2010-07-25 4.21a (PWD, AO, DPC) hacked
%Control: key (0)
%Control: author (72) initials jnrlst
%Control: editor formatted (1) identically to author
%Control: production of article title (-1) disabled
%Control: page (0) single
%Control: year (1) truncated
%Control: production of eprint (0) enabled
%
\end{document}